\documentclass[12pt]{article}
\usepackage{latexsym,amsmath,amssymb,amsbsy,graphicx}
\usepackage[cp1251]{inputenc}
\usepackage[T2A]{fontenc}
\usepackage[russian,english]{babel}
\usepackage[space]{cite} 

\begin{document}
\bigskip

{\bf

Cyclic Variations in the Solar Radiation Fluxes
at the Beginning of the 21st Century
}

\bigskip

\centerline {E.A. Bruevich, V.V. Bruevich, G.V. Yakunina}

\centerline {\it Sternberg Astronomical Institute, Lomonosov Moscow State
 University,}
\centerline {\it Universitetsky pr., 13, Moscow 119992, Russia}\

\centerline {\it e-mail:  {red-field@yandex.ru, yakunina@sai.msu.ru}
 }\

\bigskip

{\bf Abstract.} The solar activity in the current, that is, the 24-th, sunspot cycle is analyzed. Cyclic variations in the sunspot number (SSN) and radiation fluxes in various spectral ranges have been estimated in comparison with the general level of the solar radiation, which is traditionally determined by the radio emission flux $F_ {10.7} $
at a wavelength of 10.7 cm (2.8 GHz). The comparative analysis of the variations in the solar constant and
solar indices in the UV range, which are important for modeling the state of the Earth's atmosphere, in the
weak 24th cycle and strong 22nd and 23rd cycles showed relative differences in the amplitudes of variations from the minimum to the maximum of the cycle. The influence of the hysteresis effect between the activity indices and $F_ {10.7} $ in the 24-th cycle, which is considered here, makes it possible to refine the forecast of
the UV indices and solar constant depending on the quadratic regression coefficients that associate the solar indices with $F_ {10.7} $ depending on the phase of the cycle.

\bigskip

\vskip12pt
\section{Introduction}
\vskip12pt

 The analysis of solar activity is of a great practical importance for understanding the physical processes on the Sun and predicting space weather and processes in the Earth's atmosphere. Since 1650, the solar activity has been estimated by the relative sunspot number
(SSN). Regular observations of the solar radio flux at a wavelength of 10.7 cm (2.8 GHz), which is also called the $F_ {10.7} $ index, began in 1947.

The $F_ {10.7} $ index is associated with the radiation from the entire disc and is currently used for monitoring and forecasting of solar activity more often than other indices. The solar radiation
variability in the SSN and $F_ {10.7} $ observations is cyclic; the duration of the main cycle of the sunspot activity (Schwabe's-Wolf's cycle) is approximately equal to 11 years. The current, 24th, cycle is the weakest in more than 100 years.

The next solar activity minimum between cycles 24 and 25 is expected in 2018-2019.
The 24th cycle retains the tendency of the recent years associated with a noticeable decrease in the sunspot variations in the 11-year cycle. After the maximum of the 21st cycle (in approximately 1980), cycles 22-23 did not follow the even-odd sequence of the Gnevyshev-
Ohl rule. According to this rule, the even solar cycles are weaker than the consequent odd cycles.
Many NASA scientists who work on predicting solar activity believe that the 25th cycle will be similar to the 24th or weaker.

Figure 1 illustrates the series of the sunspot number, SSN, and relative sunspot areas, A, measured in millionths of visible hemisphere (mvh). The data were obtained by indirect estimates for the 1600-1850 period and by direct observations for the period
from 1850 to the present time (reliable series) [1]. In the strong cycles, the SSN maximum reaches approximately 250 and the total area A reaches 1400.

The amplitudes in the weakest cycles are four or five times smaller. It can be seen that the SSN and A series closely correlate with each other. The Maunder minimum (from the sunspot area estimates for 1645-1715) and Dalton minimum (1790-1820) are noted.

The minimum observed now is similar in shape to the Dalton minimum. It is associated, apparently, with the simultaneous occurrence of the minimums of centennial and bicentennial cyclicity at the beginning of the 21st century, which overlapped with the 11-year cycle.

In [2], the solar activity variations were numerically modelled using the sunspot number
observational data for the 1750-2050 period. The evolution equation for the sunspot number series was solved considering the mechanism of the magnetic field formation in the sunspot based on the dynamo theory.

The modelling results in [2] closely correlate with the observations and predict an extended period of low activity up to 2050, similar to the Dalton minimum: the following cycles 25 and 26 are also expected to be relatively weak. Currently, several global indices of solar activity are continuously observed to monitor the situation on the Sun and develop various predictions. The high degree of correlation of the 10.7 cm radio flux $F_ {10.7} $ with all the main activity indices implies a strong dependence of the indices on the parameters of the plasma, where these fluxes are formed, given that the regions of their formation are spatially close.

The objectives of this study are the following: 

- to study the solar activity in cycle 24 using observations in multiple bands of the spectrum;

- to determine the quadratic regression coefficients for cycle 24 and the following weak cycles to predict the values of the indices based on the general level of the solar activity with respect to the phase of the cycle.

\begin{figure}[tbh!]
\centerline{
\includegraphics[width=120mm]{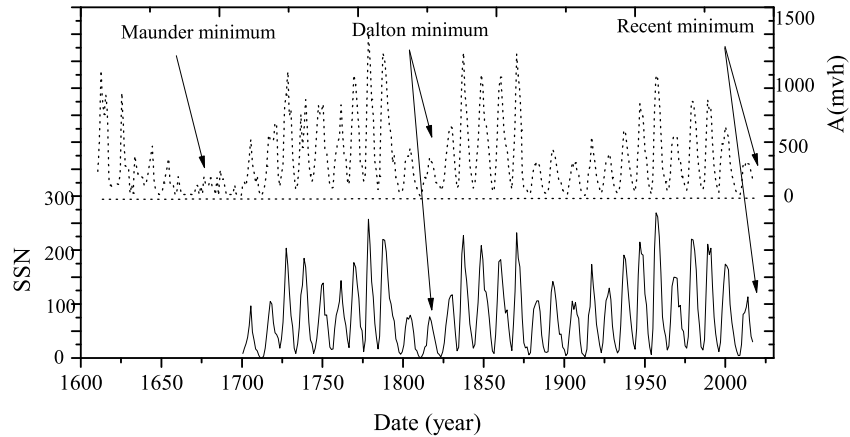}}
 \caption{The mean annual values of the relative sunspot number for the period from 1700 to 2016 (bottom) and the total sunspot areas for the period from 1600 to 2016 (top).
}
{\label{Fi:Fig1}}
\end{figure}

 The index of the relative sunspot number W (currently, the term SSN is used) was defined by Rudolf Wolf as $W = k \cdot (f + 10g)$, where f is the total number of individual spots on the disc and g is the number of spot groups. This index reflects the contribution to the solar activity not only from the sunspots themselves, but from the entire active region, mainly that occupied with faculae. The SSN index closely agrees with the more accurate contemporary indices, for example, with the radio flux $F_ {10.7} $. The observational series of these two activity indices are the longest of such indices, most of which have been observed using satellite instruments since the 1970s.

\begin{figure}[tbh!]
\centerline{
\includegraphics[width=120mm]{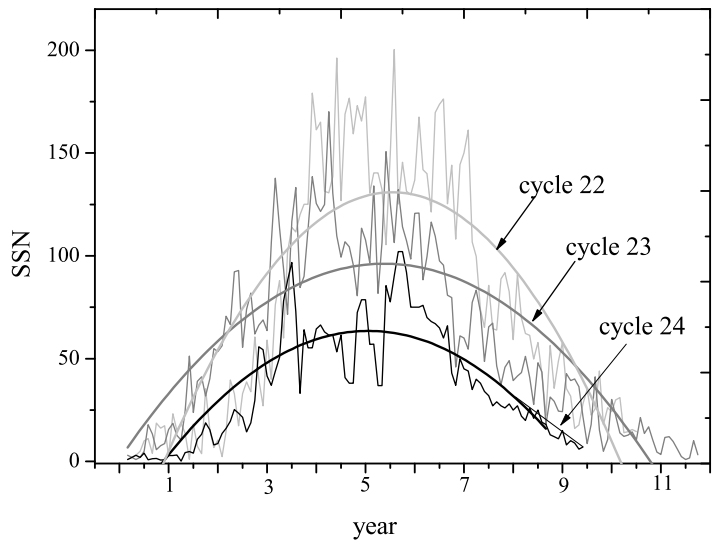}}
 \caption{The observational time series of the mean monthly
values of the relative sunspot number in cycles 22, 23, and
24. The dependence on time from the start of the cycle.}
{\label{Fi:Fig2}}
\end{figure}

\subsection{Radio Flux $F_ {10.7} $  in Cycles 22-24}

The radio flux $F_ {10.7} $, along with the relative sunspot number, is the most frequently used solar index, which characterizes the current level of solar activity. This flux is measured in solar flux units (sfu), where 1 sfu corresponds to the flux of $10^{-22}W/m^2/Hz$.
The data on $F_ {10.7} $ and the relative sunspot number are available at the archive 

http://lasp.colorado.edu/lisird/tss/noaa\_radio\_flux.html.

Figure 3 shows the mean monthly observational data for $F_ {10.7} $ in cycles 22, 23, and 24. It can be seen that cycles 22 and 23 look very similar and are almost equal in magnitude, with cycle 23 being 2 years longer. In cycle 24, the fluxes decrease by almost half in comparison with cycle 22, which agrees with the variations in sunspot numbers. The real-time $F_ {10.7} $ flux from the entire solar disc correlates quite well with the fluxes in the UV part of the solar spectrum. It can be used as a base index for predicting the fluxes in these bands of the solar spectrum in the case of gaps in satellite observations or when real-time UV observational data are required.

\subsection{Ultraviolet Flux in the MgII 280 nm Line (core/wings) in Cycles 22-24}

The UV flux from the Sun comprises a small part of the total solar radiation flux; however, it plays an important role in the formation of the upper atmosphere of the Earth. The UV and X-ray radiation fluxes serve as input parameters in all ionospheric models of the Earth's atmosphere.
The MgII index (280 nm), which characterizes the ratio between the fluxes in the center and in the wings, has been continuously observed from satellites since the 1970s. This index is a good indicator of the state of the solar chromosphere (the core of the line) and photosphere
(wings). The MgII (core/wings) index is measured in relative units.

\begin{figure}[tbh!]
\centerline{
\includegraphics[width=120mm]{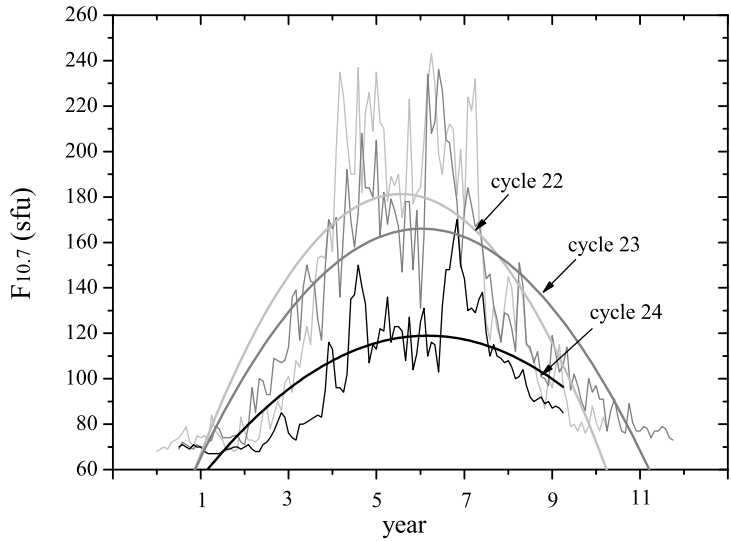}}
 \caption{The observational time series of the mean monthly values of the radio flux $F_ {10.7} $  in cycles 22, 23, and 24. The dependence on time in years from the start of the cycle.}
{\label{Fi:Fig3}}
\end{figure}

\begin{figure}[tbh!]
\centerline{
\includegraphics[width=120mm]{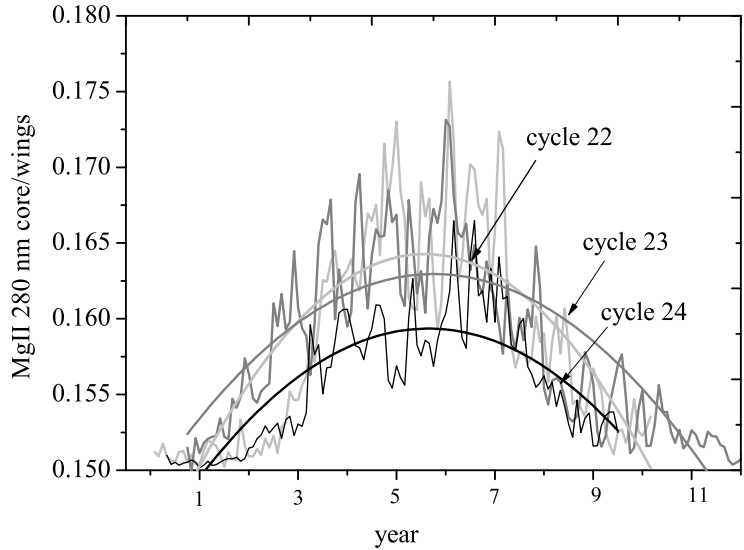}}
 \caption{The observational data of MgII 280 nm (core/wings) at the GOME (1995-2011), 
 SCIAMACHY (2002-2012), GOME-2A (2007-2017), and GOME-2B (2012-2017) satellites from the Bremen composite data archive in cycles 22, 23, and 24. The dependence on time
from the start of the cycle.}
{\label{Fi:Fig4}}
\end{figure}

Figure 4 shows the mean monthly values of the satellite observations in cycles 22-24, calibrated to a uniform absolute scale [5]. The data archive is available at http://www.iup.uni-bremen.de/UVSAT/Datasets/mgii. 

It can be seen that the maximum amplitudes in cycles 22 and 23 have almost the same value, while the maximum amplitude in cycle 24 is noticeably lower. The
MgII (core/wings) index closely correlates with $F_ {10.7} $, since these indices characterize the state of the solar activity as a whole, as well as the state of the upper chromosphere and lower corona region, where they are formed.

\begin{figure}[tbh!]
\centerline{
\includegraphics[width=130mm]{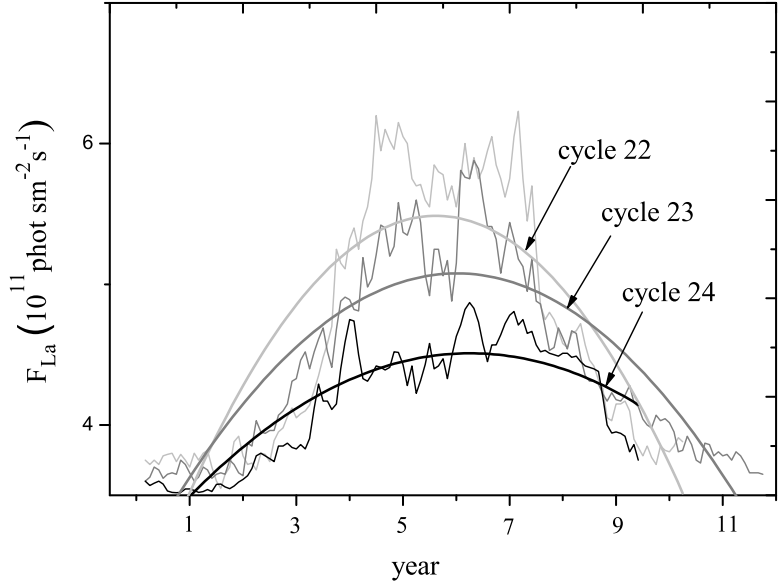}}
 \caption{The observational data of $F_ {L \alpha} $ 121.6 nm at the AEE, SME, and UARS satellites in cycles 22, 23, and 24. The dependence on time from the start of the cycle.}
{\label{Fi:Fig5}}
\end{figure}

\subsection{UV Flux in the Lyman-alpha 121.6 nm Line in Cycles 22-24}

\begin{figure}[tbh!]
\centerline{
\includegraphics[width=130mm]{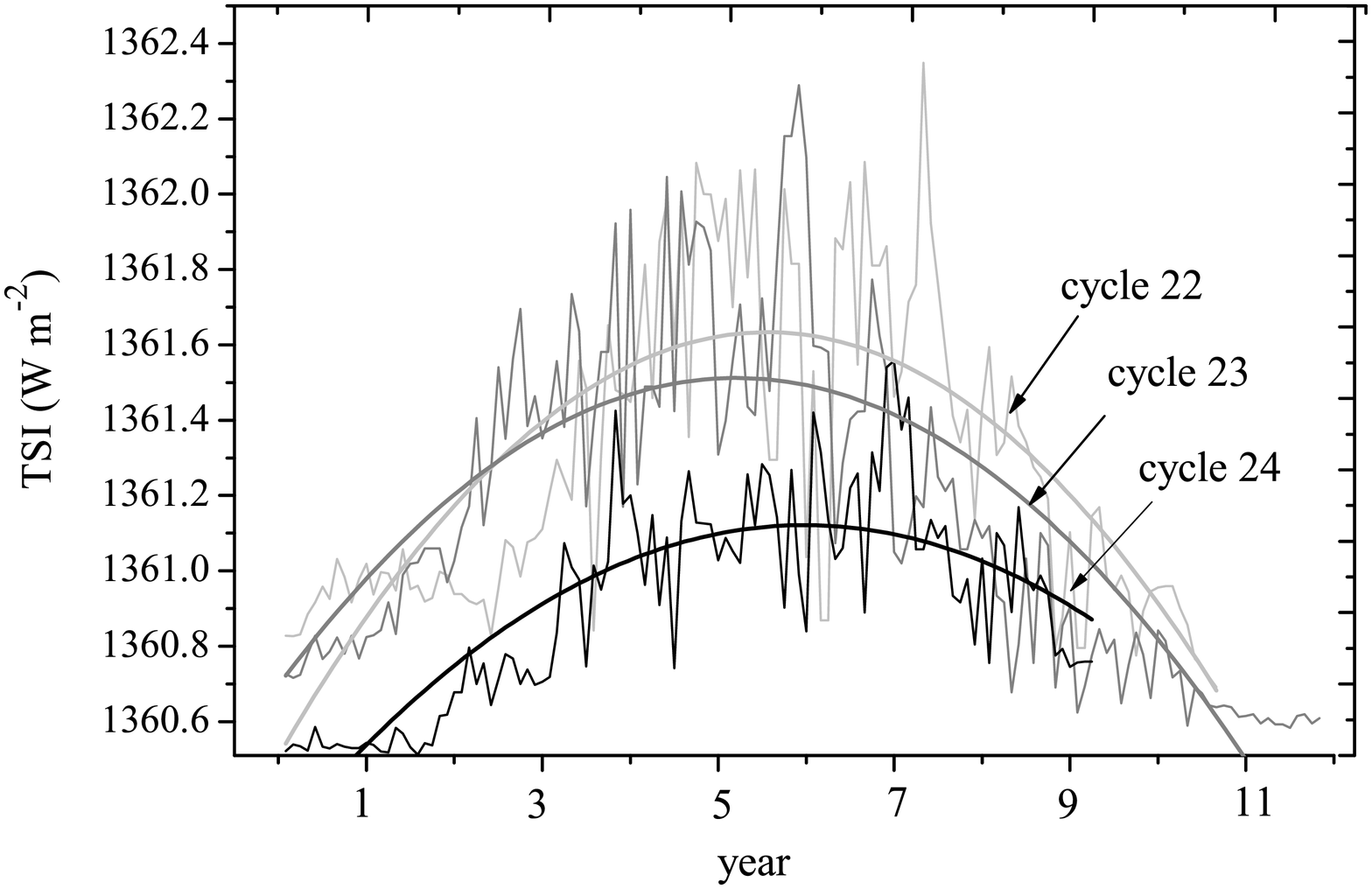}}
 \caption{The solar constant (TSI) according to the observations at the Nimbus 7 and ACRIM 1-3 satellites from the TSI PMOD composite data archive in cycles 22, 23, and 24. The dependence on time from the start of the cycle 24. The dependence on time from the start of the cycle.}
{\label{Fi:Fig6}}
\end{figure}

\begin{figure}[tbh!]
\centerline{
\includegraphics[width=110mm]{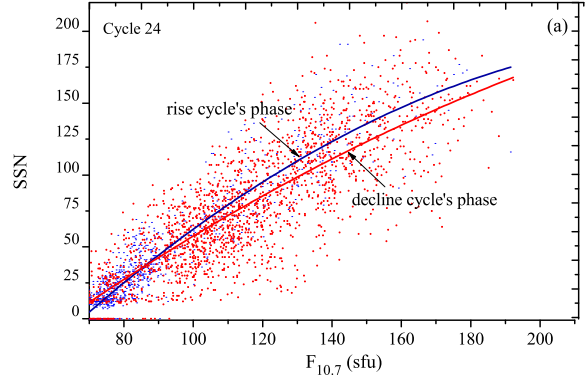}}
 \caption{The hysteresis effect of the sunspot number daily values as a function of $F_ {10.7} $.}
{\label{Fi:Fig7}}
\end{figure}

\begin{figure}[tbh!]
\centerline{
\includegraphics[width=110mm]{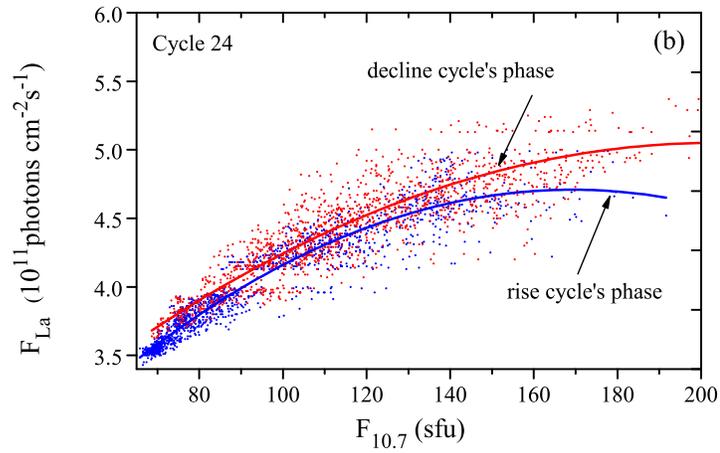}}
 \caption{the hysteresis effect of the Lyman alpha
flux (121.6 nm) daily values as a function of $F_ {10.7} $.}
{\label{Fi:Fig8}}
\end{figure}

\begin{figure}[tbh!]
\centerline{
\includegraphics[width=110mm]{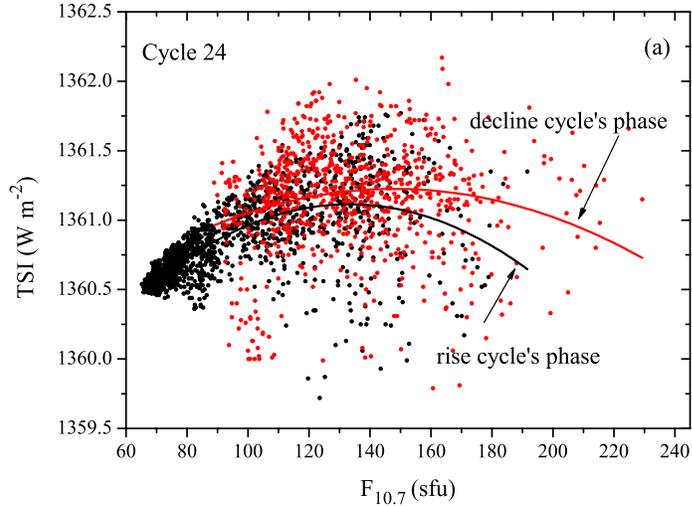}}
 \caption{The hysteresis effect of the solar constant daily values as a function of  $F_ {10.7} $.}
{\label{Fi:Fig9}}
\end{figure}

\begin{figure}[tbh!]
\centerline{
\includegraphics[width=110mm]{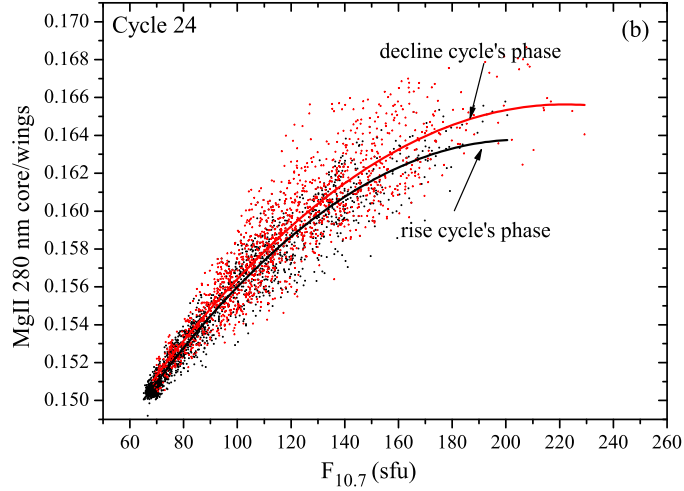}}
 \caption{the hysteresis effect of the MgII nm
(core/wings) daily values as a function of $F_ {10.7} $.}
{\label{Fi:Fig10}}
\end{figure}

\begin{figure}[tbh!]
\centerline{
\includegraphics[width=110mm]{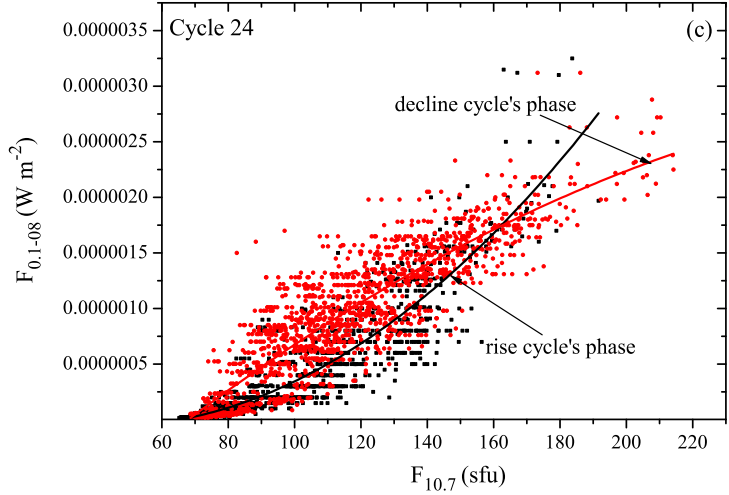}}
 \caption{the hysteresis effect of the $F_ {0.1-0.2} $ daily values as a function of $F_ {10.7} $.}
{\label{Fi:Fig11}}
\end{figure}

\begin{table}
\caption{Quadratic regression coefficients for the daily values of the solar activity indices (AI) in cycle 22 in the phases of
rise (R) and decline (D)}
\begin{center}
\begin{tabular}{clclclclclclcl}

\hline
              $ AI\leftrightarrow F_{10.7}$       &   ~~ A  &   C1  &
              ~~C2 &
  $ \sigma A$ &   $\sigma C1 $   &     $\sigma C2$    \\
\hline

  $ SSN\leftrightarrow F_{10.7}, R $  &   -169.54   & 2.88    &   -0.0057 &
5.32    & 0.07   & 5.18E-4 \\
\hline

$ SSN\leftrightarrow F_{10.7}, D $  &   -115.55   & 2.01    &
-0.0028 &
7.71    & 0.08   &7.47E-4     \\
\hline

 $ TSI\leftrightarrow F_{10.7}, R$ &  1358.68       & 0.037    & -1.37E-4     &
  0.071 & 0.0014  & 6.71E-6        \\
\hline

 $ TSI\leftrightarrow F_{10.7}, D$   &  1359.58    & 0.022     &   -7.51E-5   &
  0.301   & 0.0043  &   1.49E-5        \\
\hline

 $F_{L \alpha} \leftrightarrow F_{10.7}, R$&  1.43    & 0.039    &   -1.15E-4   &
1.43   & 9.38E-4 &  4.32E-6    \\
\hline

 $ F_{L \alpha} \leftrightarrow F_{10.7}, D$ &  1.94    & 0.031    &  -7.48E-5    &
 0.062   & 9.87E-4 & 3.79E-6        \\
\hline

$MgII \leftrightarrow F_{10.7}, R$ &  0.134    & 2.85E-4    &  -6.91E-7   &
2.52E-4   & 4.9E-6   & 2.23E-8        \\
\hline

$MgII \leftrightarrow F_{10.7}, D$ &  0.135   & 2.75E-4   &  -6.19E-7    &
 5.33E-4   & 8.65E-6 & 3.39E-8       \\
\hline
 
  $F_{0.1-0.8} \leftrightarrow F_{10.7}, R$&  2.03E-7    & -1.15E-8    &   1.3E-10   &
1.1E-8   & 1.2E-9 &  5.81E-12    \\
\hline

 $ F_{0.1-0.8} \leftrightarrow F_{10.7}, D$ &  -1.7E-6    & 2.8E-8    &  -4.1E-11    &
 0.9E-7   & 1.1E-9 & 3.19E-12        \\
\hline

\end{tabular}
\end{center}
\end{table}

The radiation flux in the UV Lyman-alpha line of hydrogen ($F_ {L \alpha}$) from the entire solar disc is an important indicator that characterizes the state of the chromosphere
and the transition region [6]. $F_{L \alpha} $ is measured in $10^ {11} phot /cm^2 /s $.

Figure 5 shows the mean monthly data of the satellite observations from 1976 to the present time calibrated to a uniform absolute scale in the Laboratory for Atmospheric and Space Physics at the University of Colorado (Lyman-Alpha Composite). The archive of these data is available at 

http://lasp.colorado.edu/lisird/tss/composite\_lyman\_alpha.html. It can be seen that the maximum amplitudes in cycles 22 and 23 differ only slightly, while the maximum amplitude in cycle 24 is significantly lower.

The close correlation between $F_ {L \alpha} $ and the 10.7-cm flux was used in the Lyman-Alpha Composite when there were no direct satellite observations; $F_ {L \alpha} $ was calculated according to its current correlative relationships with $F_ {10.7} $.

\subsection{Solar Constant (Total Solar Irradiance, TSI) in Cycles 22, 23, and 24}
The solar constant is the total power of solar radiation that passes through a unit area orthogonal to the flux at a distance of 1 AU from the Sun outside the Earth's atmosphere. Its value is affected by two factors: the distance between the Earth and the Sun, which varies over the year due to the ellipticity of the Earth's orbit, and variations in the solar activity. The solar-cycle variations are associated with the changes in the radiation flux, which, in turn, is influenced by the changes in the number and total area of sunspots, as well as the total area and relative brightness of faculae. The radiation flux varies most strongly in the ultraviolet, X-ray, and radio ranges. There are two major series of the total solar irradiance observations (TSI Composite): ACRIM and PMOD. From 1978 to 2008, the ACRIM Composite demonstrated a weak positive trend in time, while the PMOD Composite demonstrated a weak negative trend.

Figure 6 shows the mean monthly data of the satellite observations in cycles 22-24 calibrated to a uniform absolute scale (TSI PMOD Composite) [7, 8]. As in the case of the satellite observations in the UV part of the spectrum, there is no single satellite that has continuously measured the TSI over the last 30 years. The comparative analysis of contemporary series scaled to a uniform standard using various models (ACRIM Composite and PMOD composite) has shown that the latest series, which uses the SATIRE-S calibration model, reflects the TSI variations most accurately [9].

It can be seen that the solar constant variations in cycles 22 and 23 are almost the same, while the fluxes in cycle 24 are significantly weaker.

\section{The hysteresis effect of the solar indices in cycle 24:
prediction of the flux values}

According to the recent studies, in particular [2], the following cycle 25, which is expected to begin in 2018-2019, will be approximately equal to cycle 24 or slightly weaker. In [10], based on the analysis of the hysteresis effect (manifested as an ambiguous relationship between the solar radiation fluxes in the rising and declining phases of the cycle), patterns were found that allowed prediction of various activity indices for cycles 22 and 23 from the value of the $F_ {10.7} $ flux. Since there are considerable differences (with the significance level $\alpha = 0.05)$ in the relationships between the indices and $F_ {10.7} $ in the rising and declining phases of the cycle, it is important to study the regression relationships for these phases separately, to increase the quality of prediction of the indices. Studying the hysteresis effect in the solar indices in cycle 24 can be useful for predicting the values of the radiation fluxes in cycle 25, which is expected to be similar to cycle 24 in strength and duration. The quadratic regression coefficients that describe the relationships between the activity indices and $F_ {10.7} $ flux in the rising and declining phases of a weak cycle, such as cycle 24, are listed in Table 1. The quadratic regression coefficients for predicting the radiation fluxes in a strong cycle (such as 22) considering the hysteresis effect were presented in [10].

Figure 7 shows the variations in the daily sunspot number depending on $F_ {10.7} $ in cycle 24; the solid lines show the polynomial regression curves in the rising and declining phases of the cycle, calculated for the daily observational data. The hysteresis effect of the solar indices is studied using second-order polynomials. Using polynomials of a higher order is not reasonable, since the corresponding members of the regression equations are negligibly small. The residual sums of the squares (RSS), that is, a measure of the discrepancies between the observational data and the model (regression line) were estimated. A low RSS value indicates close adherence of the model to the data. In our case, when the dependence of the activity indices on $F_ {10.7} $ is described using the second-order polynomials, the RSS values are minimal. 

It can be seen that there is an ambiguous connection between the indices in the rising and declining phases of the cycle, which is associated with the delay between the times of the minimums of the cycle for these indices [10, 11]. The maximum hysteresis value (the highest relative deviation in the cycle) is approximately 10-15 \%, which is significant for predictions of the solar indices. 

Figure 8 shows the data of the daily $F_ {L \alpha}$ values depending on $F_ {10.7} $ in cycle 24. The solid lines show the regression curves in the rising and declining phases calculated for the daily observational data. Table 1 lists the regression coefficients for the daily values of the solar activity indices separately for the rising and declining phase, in accordance with quadratic regression equation (1). It can be seen that the hysteresis value (the highest relative deviation in the cycle) in cycle 24 reaches approximately 10-12 \%. 

Figure 9 shows the daily TSI values from the PMOD composite TSI observations as a function of $F_ {10.7} $ in cycle 24. The regression curves are shown for the rising and declining phases in accordance with quadratic regression equation (1). It can be seen that the hysteresis effect reaches approximately 15 \% of the maximum amplitude of the TSI variation in the cycle.

Figure 10 shows the daily values of the MgII 280 nm (core/wings) index from the MgII Bremen
composite data as a function of $F_ {10.7} $ in cycle 24. The hysteresis value in cycle 24 reached approximately 7-10 \% of the maximum amplitude of the MgII 280 nm (c/w) variations.

Figure 11 shows the daily values of the X-ray nonlared fluxes in the 0.1-0.8 nm ($F_ {0.1 – 0.8} $) range using the observational data from the GOES-15 satellite [12]. The hysteresis magnitude is approximately 10 \% from the maximum $F_ {0.1 – 0.8} $ amplitude in cycle 24.

The quadratic regressions that are shown in Figs. 7, 8, and 9-11 for the activity indices are described by Eq. (1) for both the rising and declining phases:

     $$      AI = A + C1 \cdot F_ {10.7} + C2 \cdot F_ {10.7} ^2    ~~~~~~~~~~~~          (1) $$                                 
     
where A, C1, and C2 are the quadratic regression coefficients listed in Table 1. Using these quadratic regression coefficients, it is possible to predict the daily values of the solar activity indices from the real-time $F_ {10.7} $ flux. Taking the difference between the regression coefficients for the rising and declining phases of the activity cycle into account allows the accuracy of prediction of the indices to be improved.

\vskip12pt
\section{Conclusions}
\vskip12pt

1.	The analysis of solar cycle 24, which is the weakest cycle in the last 100 years, has shown that the relative differences in the amplitudes of the solar activity index variations from the minimum to the maximum of the cycle significantly change when moving from cycles 22 and 23 to cycle 24. The sunspot number and the $F_ {L \alpha} $ flux noticeably decreased in cycle 23 in comparison with cycle 22. All other activity indices have approximately the same amplitudes in cycles 22 and 23 and decrease by half in cycle 24.

2. The study of the hysteresis effect between the activity indices and $F_ {10.7} $ in cycle 24 has shown that
there are significant differences in the regression coefficients in the rising and declining phases of this cycle. The regression coefficients for different phases of the cycle are listed in Table 1. Taking the influence of the hysteresis effect in cycle 24 into account will make it possible to improve the prediction of the variations in the UV indices, the solar constant, and the X-ray non-flared flux $F_ {0.1 – 0.8} $ during weak cycles.

\end{document}